\documentclass[manuscript]{aastex}

\usepackage{graphicx}
\usepackage{color}
\usepackage{soul} 

\shorttitle{Luminous star in M33}
\shortauthors{Miko{\l}ajewska et al.}

\begin{document}


\title{Characterization of the Most Luminous Star in M33:\\ A Super Symbiotic Binary}
\author{Joanna Miko{\l}ajewska\altaffilmark{1}}
\affil{N. Copernicus Astronomical Center, Bartycka 18, PL 00--716 Warsaw, Poland}
\email{mikolaj@camk.edu.pl}

\author{Nelson Caldwell\altaffilmark{2}}
\affil{Harvard-Smithsonian Center for Astrophysics, 60 Garden Street, Cambridge, MA 02138, USA}

\author{Michael M. Shara\altaffilmark{3}}
\affil{Department of Astrophysics, American Museum of Natural History, Central Park West at 79th Street, New York, NY 10024, USA}

\and

\author{Krystian I{\l}kiewicz\altaffilmark{1,4}}
\affil{N. Copernicus Astronomical Center, Bartycka 18, PL 00--716 Warsaw, Poland}

\altaffiltext{4}{Warsaw University Observatory, Al. Ujazdowskie 4, 00-478 Warszawa, Poland}



\begin{abstract}
We present the first spectrum of the most luminous infrared star in M33, and use it to demonstrate that the object is almost certainly 
a binary composed of a massive O star and a dust-enshrouded Red Hypergiant.
This is the most luminous symbiotic binary ever discovered.
Its radial velocity is an excellent match to that of the hydrogen gas in the disk of M33, supporting our interpretation that it is a very young
and massive binary star. 
\end{abstract}


\section{Introduction and Motivation}

The existence of core-collapse supernovae with dust-enshrouded progenitors  \citep{pri08, tho09} or recent mass loss episodes \citep{gal07, smi08} is a
strong motivator to survey nearby galaxies for such objects. Understanding these luminous, dusty stars is essential to any theory of mass loss from massive stars,
an important and unsolved problem in stellar evolution theory. Surveys and studies of such stars have been reported for the Large and Small Magellanic Clouds \citep{bon09, bon10}, and M33, NGC 300, M81 and NGC 6946 \citep{tho09, kha10}. The \citet{tho09} and \citet{kha10} studies demonstrated that the extremely red (optically thick even at 3.6  $\mu$m) progenitors of SN2008S and a very luminous NGC 300 transient are very rare, with only of order $\sim$\ 1 existing at any instant in a galaxy. This implies that these objects are extremely short lived phases in the lives of some massive stars.

\citet{kha11} recently identified and studied the most luminous mid-infrared source in M33 (which they named ``Object X"). 
Khan et al. measured i band variability of Object X of up to 0.4 magnitudes on a timescale of one year. They deduced that 
``The correlated short-term variability of ~0.4 mag (fractional variability of $\sim$45\,\%) definitively indicates that it is a single stellar object rather than multiple objects blended together."
We note that this amplitude of variability does not preclude Object X from being a binary, and we will show this to be the case below.

Using multiple surveys' photometry from the U band to 24  $\mu$m, \citet{kha11} considered several models for the star, and suggested that a cool hypergiant star, emitting $\sim 10^{5.6} L_{\sun}$, best matched the available data. They also predicted that the star might be emerging from the cocoon of dust in which it is embedded, leading to a future brightening in near-IR and possibly even visible bands. The most challenging observation for the Khan et al model is the presence of strong H$\alpha$ emission in Object X, evident in comparisons of narrowband and broadband images of the star from the Local Group Galaxy Survey \citep{mas06}. 

\citet{kha11} encouraged further study of this extremely luminous star, motivating us to acquire its optical spectrum. As we describe below, the spectrum suggests that Object X is a massive, luminous binary with characteristics reminiscent of Symbiotic Stars (SySt). In Sect.\,\ref{obs} we describe the data and their reductions. 
We demonstrate that Object X must be a binary and characterize the underlying stars in Sect.\,\ref{discussion}. A brief summary of our results is found in Sect.\,\ref{conclusions}.

\section{Observations and Data Reduction}\label{obs}

The spectra of Object X were obtained with the Hectospec multi-fiber positioner and spectrograph on the 6.5m MMT telescope \citep{fab05}, 
as described in \citet{cal09}. The Hectospec 270 gpm grating was used and provided spectral coverage from roughly $3700-9200${\AA} at a resolution of 
$\sim5${\AA}.  The observations were carried out on the nights of 22 and 23 September 2014.
We used the data reduction methodology described by \citet{cal09}. De-biasing and flat fielding of each frame were followed by extraction of spectra  and wavelength calibration. ``Blank sky" fibers spectra from the same exposure
were averaged to enable sky subtraction. The instrument response and flux calibration 
were enabled via the spectra of standard stars taken over the course of our observing run. Relative line flux ratios accuracy is ensured by application of careful relative flux correction from the standard stars.
The total exposure time was 1800 s.

Archival images of Object X were retrieved from the Local Group Galaxies Survey (LGGS, Massey et al. 2006, 2007), and from the Canada-France-Hawaii Telescope (CFHT) archives.
We repeated the measurements of Object X on the LGGS images that have been done by \citet{kha11}. Following their method we used
DAOPHOT/ALLSTAR \citep{stetson1992} for photometry and bright sources in the LGGS catalog for calibration. 
In the case of the frame in the Halpha filter the emission from the HII region introduced a nonlinear background near Object X. 
This background introduced significant non-repeatability of the measured brightness of Object X as we varied the DAOPHOT parameters used in Point Spread Function (PSF) fitting. 
To overcome this we fitted a low-order function to the nearby background in order to remove it. The same procedure was carried out for comparison stars. As a result we obtained a reasonably good fit of the PSF and a brightness that
depended only weakly on the DAOPHOT parameters.

\begin{figure}
\centerline{\includegraphics[width=0.45\columnwidth]{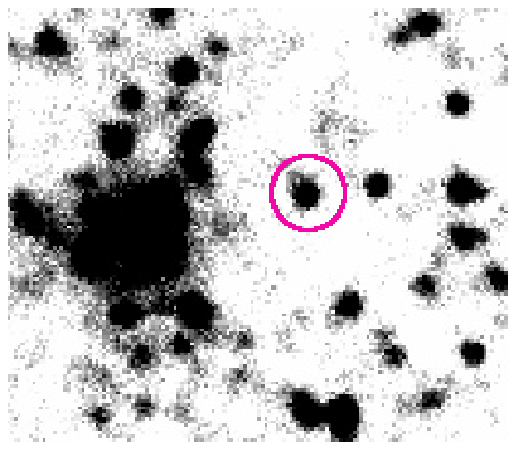}\includegraphics[width=0.45\columnwidth]{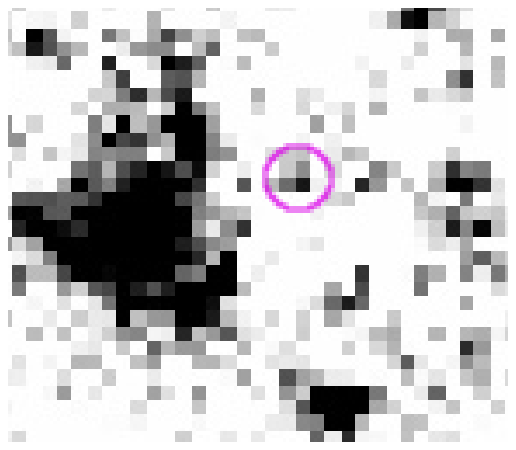}} 
\caption{I-band images of the region near Object X  (circled in the figure) obtained on Feb 18, 2001 (left) and on Oct 29, 2014 (right). Comparison with objects of similar brightness to the west and southwest (see text) demonstrate that Object X has not brightened significantly during this time.
}
\label{image}
\end{figure}

We supplemented photometry of LGGS images with observation carried out with the CFH12K camera \citep{cuill2000} on the CFHT 3.6m telescope located located at the summit of Mauna Kea, Hawaii.
The method was identical to that used for the LGGS images. The standard stars were chosen from the bright sources in the LGGS catalog that do not appear in
the \citet{hart06} list of variable sources.

New images of Object X were taken with the Tenagra II 32$\arcsec$ telescope in
Arizona. The telescope is instrumented with a SITe-based CCD  camera with a
resolution of $\sim 0.87\arcsec$ per pixel. Two 300s observations through the I filter were carried out on 29 October 2014.
The data reduction was performed using standard IRAF procedures. In these observations the TENAGRA PSF
was highly undersampled, and the relatively low S/N in the images caused the background close to Object
X to be dominated by instrument noise. Because of this we carried out aperture photometry of Object X
and nearby standard stars using DAOPHOT. We used standard stars whose brightnesses we measured in the
LGGS image.

The optical magnitude of Object X was also estimated by multipling the spectrum with the $V$ filter function of \citet{bessell1990} and adding the constant, which was calculated as the average offset between the convolution of the spectrum with the $V$ filter and the $V$  magnitudes measured on the LGGS frame for several H$\alpha$-bright stars observed simultaneously with Object X during our September observing run. 

\section{Discussion}\label{discussion} 

\subsection{Variability}

\citet{kha11} suggested that Object X may conceivably emerge from its self-obscured state and brighten over the next few decades. We find no sign of this, as evidenced by our recent I-band image (Figure\,\ref{image}). In particular, $I=20.4\pm0.6$ for Object X derived from this 2014 image when compared to $I=19.99\pm0.02$ reported by \citet{kha11} indicates that the source is definitely not brightening.

Similarly, we detected Object X at $V=22.88\pm0.05$ in the CFHK12 image taken on 1999 Oct 30, 2 yrs before the LGGS observation, and on two other CFHK12 images at $V=22.72\pm0.07$ (2001 Feb 18) and $V=22.70\pm 0.06$ (2001 Oct 15). These magnitudes are in very good agreement with our  measurements of $V=22.86\pm0.10$ derived from the LGGS image, and $V=22.8\pm 0.2$ estimated from our spectrum.
The difference between our LGGS $V$ magnitude and $V=23.15\pm0.13$ measured by \citet{kha11} most probably originates in our very careful subtraction of the background, which is highly variable in that region due to the nearby HII region. 
We conclude that the optical brightness of the Object X has remained stable at $V=22.8\pm 0.1$ mag over at least the last 15 years.
Similarly, the H$\alpha$ magnitude measured on the LGGS image, $m(\rm 
H\alpha) = 20.4\pm0.2$ reasonably agrees with the value of about 20.7 estimated from the spectrum (described in the next subsection) using the mean flux in a 50\,\AA\ interval centered on H$\alpha$. 
It remains to be seen whether Khan et al.'s prediction of a significant brightening of Object X will, in fact, occur in the coming decades.

\subsection{The Hot Star and the Ionised Region}

\begin{figure}
\centerline{\includegraphics[width=0.98\columnwidth]{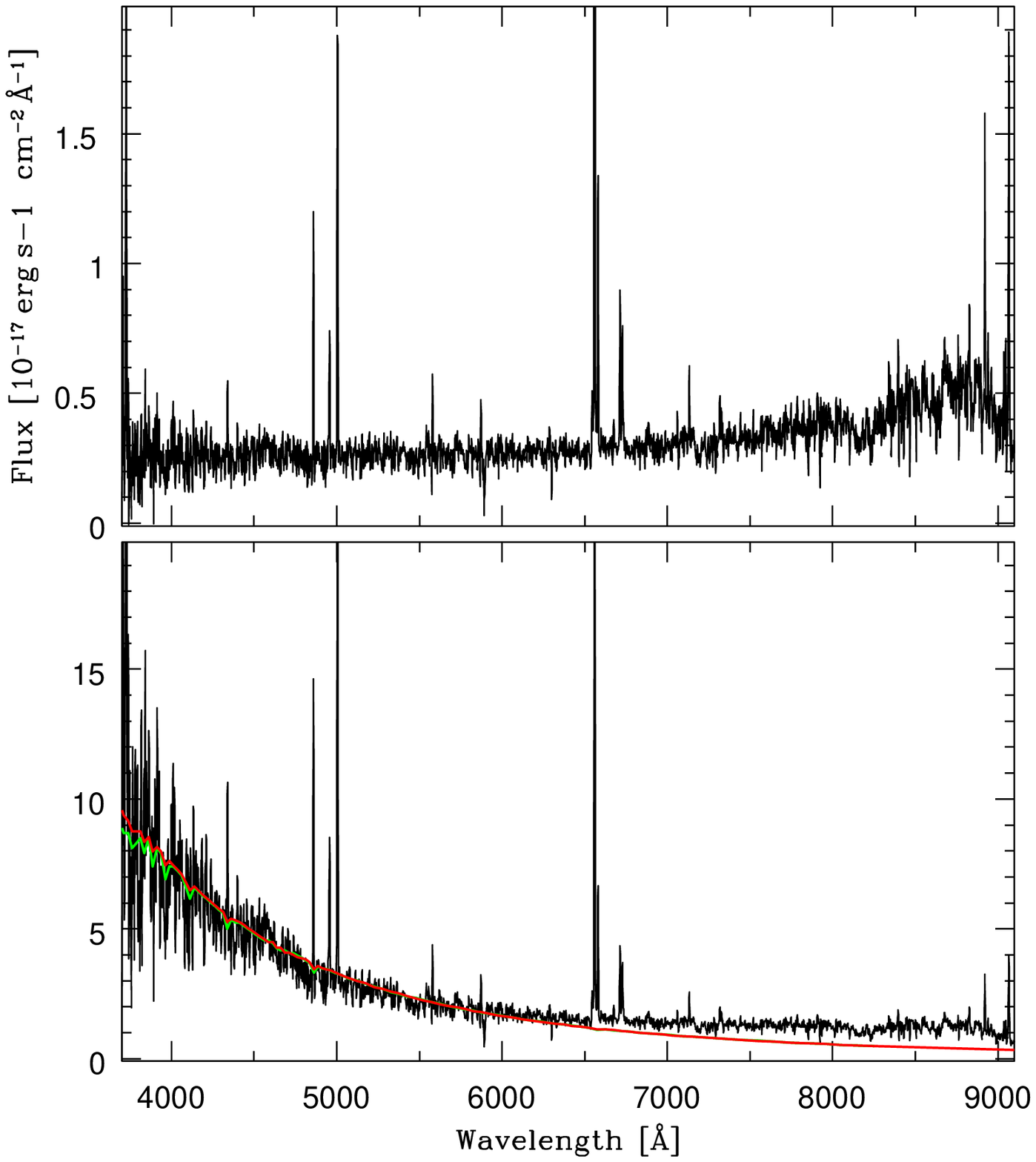}}
\caption{(top) The spectrum of Object X observed in September 2014. (bottom) The same spectrum dereddened with $E_{\rm B-V}=0.92$ (see text). The green and red lines represent the Kurucz model atmosphere spectra with $T_{\rm eff}= 35\,000$ K, and 45\,000 K, respectively.}
\label{spectrum}
\end{figure}

The optical spectrum of Object X observed in September 2014, and normalized to $V=22.8$ is shown in Figure\,\ref{spectrum}, and the emission line fluxes and other measurements from that spectrum are listed in Table 1. 
 
The outstanding feature of this spectrum is the presence of strong, high ionization potential (IP) emission lines, most notably [\ion{O}{3}]\,5007.  The inescapable conclusion is that {\it Object X has a hot, luminous component}, in addition to the cool, infrared-bright component discussed by \citet{kha11}. 

The forbidden [\ion{O}{2}],[\ion{O}{3}],[\ion{N}{2}], and[\ion{S}{2}] to Balmer emission line ratios are consistent with a photoionized \ion{H}{2} region rather than a shock-excited region \citep{canto} or diffuse interstellar gas (DIG; e.g. Hoopes \& Walterbos 2003). 
The average radial velocity from 14 unblended emission lines listed in Table\,1 is --$113\pm5\,\rm km\,s^{-1}$, and there is no trend in the velocity nor in the line widths with the IP (as expected for an \mbox{{H}\,{\sc ii}} region) which means that the lines do not originate in the hot star wind (not surprising given the line widths -- the
FWHMs correspond to the instrumental value).

The Balmer line ratios, H$\alpha$:H$\beta$:H$\gamma$=6.63:1:0.30,  are consistent with case B recombination and a reddening $E_{\rm B-V}$=0.92$\pm0.02$, adopting the mean extinction law of \citet{CCM}, and  $R_{\rm V}=A(V)/E(B-V)=2.5$ similar to that in the Large Magellanic Cloud \citep{misselt1999}. 
\begin{table}
\begin{center}
  \caption{Emission line fluxes of Object X in M33}\label{tabl}
  \begin{tabular}{|ccc|}
 \tableline
$\lambda_{\rm obs}$ & ID & 100\,$F(\lambda)/F({\rm H\beta})$ \\
\tableline
3726.9 & [\ion{O}{2}]\,3728\tablenotemark{2} & 225\\
4338.9 & H$_\gamma$ & 30\\
4859.5 & H$_\beta$	& 100\\
4956.7 & [\ion{O}{3}]\,4958.9 & 56\\
5004.8 & [\ion{O}{3}]\,5006.8 & 165\\
5873.1 &\ion{He}{1}\,5875.7 & 24\\
6546.0 & [\ion{N}{2}]\,6548.1 & 27\\
6560.2 & H$_\alpha$ & 663 \\
6580.8 & [\ion{N}{2}]\,6583.5 & 99\\
6672.8 &\ion{He}{1}\,6678.2 & 13\\
6714.0 & [\ion{S}{2}]\,6716.4 & 65 \\
6728.1 & [\ion{S}{2}]\,6730.8 & 50 \\
7062.0 &\ion{He}{1}\,7065.2 & 7\\
7134.0 & [\ion{Ar}{3}]\,7135.8  & 26\\
7316.9 &[\ion{O}{2}]\,7320\tablenotemark{3} & 19 \\
7327.1 & [\ion{O}{2}]\,7330\tablenotemark{3} & 18\\
9065.8 & [\ion{S}{3}]\,9068.6 & 149 \\
\tableline
\multicolumn{2}{|c}{$\log F({\rm H\beta})$\tablenotemark{1}} & -16.25 \\
\tableline
\multicolumn{2}{|c}{$v_{\rm r}$} & --$113\pm5\, \rm km\,s^{-1}$ \\
\tableline
\end{tabular}
\tablenotetext{1}{in units of $\rm erg\,s^{-1}\,cm^{-2}$}
\tablenotetext{2}{blend of 2 components}
\tablenotetext{3}{blend of 3 components}
\end{center}
\end{table}

\citet{mur94} demonstrated  that the temperatures $T_{\rm h}$ of the hot components in symbiotic binary stars can be estimated from a simple formula,
$T_{\rm h}[1000\,{\rm K}]=\chi_{\rm max}[\rm eV]$, where $\chi_{\rm max}$ is the highest observed ionization potential. The accuracy of their method is about $\pm 10\,\%$.
The highest ionization stage observed in our object is that of O$^{+2}$, which suggests  a  lower limit to the ionization temperature $T_{\rm h}$ of 35\,000 K.
\citet{kal78} found that in low-excitation optically thick nebulae the [\ion{O}{3}]/H$\beta$ intensity ratio correlates well enough with the temperature of the hot ionizing star, $T_{\rm h}$, to be used alone as a good indicator of the calculated blackbody temperature. In particular, for nebulae for which the \ion{He}{2} 4686 line is weak or absent, 
and for which $T_{\rm h}$ is no more than about 68\,000 K, $\log T_{\rm h} = 4.426+4.827 \times 10^{-4} F(5007/{\rm H\beta}) -1.374 \times 10^{-7} F^2(5007/{\rm H\beta})$.
The relative  emission line flux for our object, F(5007/$H\beta$)=159 yields $T_{\rm h}=31200$ K, which corresponds to a B0 star, providing an independent  lower limit for the hot component temperature.

On the other hand, the presence of \ion{He}{1} emission lines indicates the presence of a star of spectral type O7 or hotter
spectrum.
Assuming that case B recombination applies for both the Balmer and \ion{He}{1} emission lines, the de-reddened $F(\mbox{He}{\sc i} 5876)/F({\rm H\beta})$=0.18 requires a blackbody temperature of the ionizing source of 45\,000 K, which provides an upper limit for $T_{\rm h}$.

In Fig.\,\ref{spectrum} we also show the spectrum de-reddened with $E_{\rm B-V}=0.92$. Superposed on this dereddened spectrum are \citet{kurucz}
model atmosphere spectra with $T_{\rm eff}= 35\,000$ K, and 45\,000 K, respectively, and scaled to $V=22.8$. This clearly demonstrates that the optical continuum is dominated by the hot O/B star with the reddening practically identical with that derived from the Balmer decrement.
We thus conclude that this hot star is responsible for the ionization of the \mbox{{H}{\sc ii}} region coinciding with Object X.

Our average $V=22.8$, $A_{\rm V}=2.3$ combined with  the true distance modulus $m-M=24.62$ \citep{gie13} results in  the absolute magnitude of the hot star  $M_{\rm V} = -4.2 \pm0.5$ (where the error includes the uncertainties in the reddening, $V$ mag and distance estimates). 
Adopting the bolometric correction, $BC \sim -3.4 \div -4.1$ \citep{kurucz}, corresponding to the $T_{\rm h}$ range indicated by the emission line fluxes, the bolometric magnitude of the hot star is then $M_{\rm bol} \sim -7.6 \div -8.3$.
The number of photons capable of ionizing hydrogen, emitted by such a star, is sufficient to account for the observed intensities of the Balmer lines.

\subsection{The Evolutionary State of the Binary Components of Object X}
 
The hot star luminosity ($\log L \sim 4.9 \div 5.2$) and effective temperature ($\log T_{\rm h} \sim 4.51$--4.65) suggest that it is still on or close to the ZAMS.
The evolutionary tracks and isochrones for massive stars (e.g. \citealt{Brott2011}) suggest that the hot star mass is about 20--30\,$\rm M_{\sun}$, and that its age is at most 2--3 Myr.
\citet{kha11} demonstrated that the infrared SED of Object X can be interpreted in terms of a dust-obscured, massive, $\ga 30\, \rm M_{\sun}$, evolved star, presumably a yellow/red hypergiant. Their finding is consistent with our result since the more evolved component should be also the more massive one.

The average radial velocity of Object X derived from emission lines, $v_{\rm r}=-113\, \rm km\,s^{-1}$ is redshifted with respect to the mean radial velocity of M33.
It is also in excellent agreement with the mean velocity of hydrogen gas at the galactic position of Object X, viz. 
 $-105 \div -112\, \rm km\,s^{-1}$ \citep{rogstad}. This means that this velocity shift is due to the galactic rotation, and the system lies close to the galactic disk plane.
The maximum reddening internal to M33 is of $E_{\rm B-V} \sim 1.2$ mag \citep{kha11}. Thus the reddening of $E_{\rm B-V}=0.92$ we estimated for Object X is likely of interstellar origin.

\subsection{Object X and D-type Symbiotic Stars }

The binary nature of Object X is not surprising given that  
the binarity and
multiplicity fraction in high-mass stars seems to be of the order of
100\,\% 
(e.g \citealt{mason2009}).
What is particularly interesting in the case of Object X is
the simultaneous presence of a hot luminous star and the very luminous, and presumably evolved component embedded in an optically thick dust shell. Such a binary configuration is reminiscent in some ways of the well known D-type symbiotic stars.

Symbiotic binary stars (SySt) are cool, evolved giants paired with a luminous, hot companion. That companion is often, but not always a white dwarf star.  The ionized, dense nebula that surrounds the binary system
emanates from the red giant. Two subclasses of SySt arise from the different kinds of red giants in these binaries. S - type (for ``stellar") SySts' cool stars are ordinary red giants. The orbital periods of S -type SySt are typically a few years  \citep{mik12}. D - type (for ``dusty") SySts' cool stars are Mira variables embedded in warm dust \citep{whi87}; their orbital periods are decades or longer (e.g. \citealt{gro09}).

Both in D-type SySt, and Object X, the hot star seems to be located outside the dust shell. 
In Object X, the reddening towards the emission line region and the hot star is presumably due to foreground reddening in M33.
Similarly, in D-type SySt the reddening towards the Mira component is usually much higher than towards the nebula, and the hot continuum (e.g. \citealt{mik99}).
According to \citet{kha11} the outer radius of the Object X dust shell is of order of $5 \times 10^{16}$\,cm (their cool star model). Adopting this as a minimum binary separation, the binary period must be $\ga$\,30\,000 yrs, with a total binary mass of $\ga 50\, \rm M_{\sun}$. 
There is, however, no evidence for any binary interaction in Object X 
(though even if there is accretion in the binary it will contribute little to the observed luminosity).
There is also no significant density gradient in Object X. 
In particular, the density sensitive ratio of [\ion{S}{2}]\,6716/6731=1.3   indicates $n_{\rm e} \approx 100\, \rm cm^{-3}$, which is typical for an \ion{H}{2} region but 1--2 orders of mag lower than the densities in the outermost regions of D-type SySt nebulae. 
There are also no detectable [\ion{O}{3}], [\ion{S}{3}], and [\ion{N}{2}] auroral lines or any other signatures of higher, $n_{\rm e} \sim 10^5$--$10^6 \rm cm^{-3}$,  density regions in Object X. Such high-density regions are observed in D-type SySt, which also show a relationship between nebular density and temperature, and IP; both $n_{\rm e}$ and $T_{\rm e}$ increase with IP (e.g. \citealt{sch90}).

Both  Object X and D-type SySt during a high dust-obscuration phase show very similar near to mid-IR colors, presumably because of similar dust temperatures.
In particular, \citet{grom2009} demonstrated that the dust-obscured O-rich symbiotic Miras are located along the line representing a simple model of a cool star (with $T_{\rm eff}=2750$) inside a warm ($\sim 800\, \rm K$) dust shell of variable optical thickness. 
Object X does NOT show any signatures of a cool star (like e.g. molecular absorption bands), however its near-IR colors locate it on this model line somewhat above the most obscured SySt, with $A_{\rm V} \sim 15$. 
Even a very luminous star, with such a high optical extinction, would not be detectable in the range covered by our spectroscopic observation.
Similarly, there is no evidence for the Mira component in optical/red spectra of dust-obscured SySt.

The most remarkable property of Object X remains its total luminosity, viz. absolute magnitude $M_{\rm K}=-11.2$. That of the most luminous D-type SySt, like e.g. RX Pup, is 10 times fainter, at $M_{\rm K}=-8.7$ \citep{mik99}. We conclude that Object X is the most luminous SySt yet discovered.

Finally, the binary composition of Object X, a cool hypergiant with an O/B star companion, bears some resemblance to the VV Cep-type stars, a small class of  massive binaries. VV Cep stars are composed of a bright red supergiant and
an early B main-sequence star, and display emission lines of \ion{H}{1} and [\ion{Fe}{2}], and occasionally also [\ion{Fe}{3}] and [\ion{O}{3}] (e.g. Cowley 1969). Although VV Cep stars do not show any evidence for the presence of dust shells, one can imagine
that if their M-type supergiant component were to be enshrouded in such a thick dust cloud, they would appear very similar to Object X. 

\section{Conclusions}\label{conclusions}

We presented and discussed the first optical spectrum of the most luminous infrared star in M33. The main conclusions are as follows.

\begin{enumerate}
\item
The star is clearly composite, with a hot, luminous O/B star component, and a cool, dust-enshrouded companion.
\item The hot star's temperature lies in the range
$35\,000 \la T_{\rm h} \la 45\,000$ K and its luminosity is $M_{\rm bol}$ = -7.6 to -8.3.
\item The emission-line spectrum and total system $M_{\rm K}$ = -11.2 establishes Object X as the most luminous symbiotic binary detected to date.
\item The radial velocity of Object X is an excellent match to that of the hydrogen gas in the disk of M33, supporting our interpretation of Object X as a very young
and massive binary star. 
\end{enumerate}


\section{Acknowledgments}

This study has been supported in part by the Polish NCN grant
DEC-2013/10/M/ST9/00086.
We gratefully acknowledge the fine support at the MMT Observatory, and
The Local Group Galaxy Survey conducted at NOAO by Phil Massey and
collaborators.
This research used the facilities of the Canadian Astronomy Data Centre operated by the National Research Council of Canada with the support of the Canadian Space Agency.
We thank Michael Schwartz and Paulo Holvorcem for support in getting recent Tenagra images of Object X, and Dave Zurek for support in accessing the LGGS archival images. We also thank Scott Kenyon for pointing out the similarity between Object X and the VV Cep stars.




\end{document}